\documentclass[aps,prb,twocolumn,reprint,superscriptaddress,nofootinbib,preprintnumbers,longbibliography]{revtex4-2}

\usepackage[english]{babel}
\usepackage[utf8]{inputenc}
\usepackage{amssymb}
\usepackage{amsmath}
\usepackage{color}
\usepackage{hyperref}
\usepackage{bbm}
\usepackage{epsfig}
\usepackage{multirow}
\usepackage[normalem]{ulem}
\usepackage[dvipsnames]{xcolor}
\usepackage{xspace}

\newcommand{\e}{\varepsilon}
\newcommand{\s}{\sigma}
\newcommand{\TK}{T_{\rm K}}
\newcommand{\VM}{V_{\rm M}}
\newcommand{\GM}{\Gamma_{\rm M}}
\newcommand{\sbar}{\bar{\sigma}}

\newcommand{\up}{\uparrow}
\newcommand{\down}{\downarrow}
\newcommand{\w}{\omega}

\newcommand{\de}{{\rm d}}

\newcommand{\Pol}{\mathcal{P}}

\newcommand{\avg}[1]{\langle {#1} \rangle }
\newcommand{\GF}[1]{\langle\!\langle #1\rangle\!\rangle}

\newcommand{\M}{\mathrm{M}}

\renewcommand{\Im}{\mathrm{Im}}

\renewcommand{\Re}{\,\mathrm{Re}\!}
\renewcommand{\Im}{\,\mathrm{Im}\!}

\newcommand{\op}[1]{\hat{#1}}
\newcommand{\fk}{\op{f}^\dagger}
\newcommand{\dk}{\op{d}^\dagger}
\newcommand{\ck}{\op{c}^\dagger}
\newcommand{\fa}{\op{f}^{}}
\newcommand{\da}{\op{d}^{}}
\newcommand{\ca}{\op{c}^{}}

\newcommand{\Sec}[1]{Sec.~\ref{sec:#1}}
\newcommand{\beq}{ \begin{equation} } 
\newcommand{\eeq}{ \end{equation} }
\newcommand{\beqa}{\begin{eqnarray}}
\newcommand{\eeqa}{\end{eqnarray}}
\newcommand{\nn}{\nonumber}
\newcommand{\es}{& = &}
\newcommand{\hc}{\mathrm{H.c.}}
\newcommand{\ie}{\textit{i.e.}}
\newcommand{\eg}{\textit{e.g.}}

\newcommand{\fig}[1]{Fig.~\ref{fig:#1}}

\newcommand{\eq}[1]{Eq.~(\ref{#1})}

\hypersetup{
    bookmarks=false,        
    colorlinks=true,        
    linkcolor=red,        
    citecolor=blue,        
    filecolor=blue,      
    urlcolor=blue
}

\begin{document}

\title{Signatures of Kondo-Majorana interplay in {\em ac} response}

\author{Krzysztof P. W{\'o}jcik}
\email{kpwojcik@ifmpan.poznan.pl}
\affiliation{Institute of Molecular Physics, Polish Academy of Sciences, 
			 Smoluchowskiego 17, 60-179 Pozna{\'n}, Poland}
\affiliation{Institute of Physics, Maria Curie-Sk\l{}odowska University, 20-031 Lublin, Poland}

 \author{Tadeusz Doma\'{n}ski}
\email{doman@kft.umcs.lublin.pl}
\affiliation{Institute of Physics, Maria Curie-Sk\l{}odowska University, 20-031 Lublin, Poland}

\author{Ireneusz Weymann}
\email{weymann@amu.edu.pl}
\affiliation{Istitute of Spintronics and Quantum Information,
			 Faculty of Physics, Adam Mickiewicz University, 61-614 Pozna{\'n}, Poland}

\date{\today}

\begin{abstract}
We analyze dynamical transport properties of a hybrid nanostructure,
comprising a correlated quantum dot embedded between the source and drain electrodes,
which are subject to an {\em ac} voltage,
focusing on signatures imprinted on the charge transport by the side-attached Majorana zero-energy mode.
The considerations are based on the Kubo formula,
for which the relevant correlation functions are determined by using
the numerical renormalization group approach,
which allows us to consider the correlation effects due to the Coulomb repulsion and their interplay with the Majorana mode
in a non-perturbative manner.
We point out universal features of the dynamical conductance,
showing up in the Kondo-Majorana regime, and differentiate them
against the conventional Kondo and Majorana systems.
In particular, we predict that the Majorana quasiparticles give rise
to universal fractional values of the {\em ac} conductance
in the well-defined frequency range below the peak at the Kondo scale.
We also show this Kondo scale to actually increase with strengthening 
the coupling to the topological superconducting wire.
\end{abstract}

\maketitle

\section{Introduction}
\label{sec:intro}

Dynamical transport properties, such as  shot noise or {\em ac} 
response, give valuable information about the charge carriers
and can provide evidence for their unique character \cite{Blanter2000Sep}. 
They are hence useful for exploring the exotic phases of matter 
\cite{Moca2018Jan,Michalek2020Jun,Franke2020,Grifoni2023}
and for uncovering the subtle fingerprints of interactions 
\cite{Ng1996Jan,Lopez1998Nov,Moca2010Jun,Moca2011Dec,Crepieux2018Mar,Stefanucci2018Jun,Wang2022Nov}.
The dynamical properties can be also studied with optical means \cite{Shahbazyan2000Jun,Giazotto2023}, 
including the quantum dots (QDs) coupled to microwave cavities \cite{Lee2011May,Deng2021Sep,Blais2021May,Bulka2022Aug},
where they provide insight into photon-assisted transport \cite{Platero2004May} 
or inelastic scattering processes \cite{Zarand2004Sep,Borda2007Jun}.
Measurements of dynamical transport properties 
have been so far reported, both for the normal \cite{Basset2012Jan,Hemingway2014Sep} 
and superconducting nanostructures \cite{Kot_etal_2020,Delagrange2021} and
they are nowadays attainable with unprecedented precision \cite{Franke2021Nov}, 
opening up the field for new exciting experiments.

In particular, the dynamical response studies could be adopted to identify unique signatures of hybrid nanostructures 
with topological superconductors (TSs), harboring the Majorana 
quasiparticles. Fluctuations of the charge currents 
through various arrangements of TSs and QDs attached 
to them have been recently studied, focusing on the shot noise 
\cite{Liu2015Feb,Jonckheere2019,Perrin2021Sep,Smirnov2022May,Gruneiro2023Jul}
and coupling to a microwave cavity 
\cite{Trif_Tserkovnyak2012,Dartiailh2017,Cottet2013,Trif2018,Ricco2022,Trif2023}.
Moreover, dynamical properties of the Majorana bound states have been analyzed by the nonequilibrium Keldysh formalism, determining the finite 
frequency emission and absorption noise to all orders in the tunneling 
amplitude through a biased junction between a normal metal and topological 
superconductor \cite{Bathellier_2019}. Peculiarities of the Majorana noise 
have been also inspected for the spinless (Kitaev) counterpart of the setup studied here, demonstrating its universal features manifested by resonances 
and antiresonances appearing at characteristic frequencies \cite{Smirnov_2019}.
Such studies provided information about less conventional signatures 
of Majorana quasiparticles, which are complementary to their zero-energy 
static features reported in the tunneling measurements \cite{Lutchyn2018May,Prada2020Rev,Flensberg2021Oct}
and quench dynamics \cite{Baranski2021,Wrzesniewski2021Mar,Wrzesniewski2021Apr}.
Vast majority of these studies, however,  focused on spinless theories,
neglecting the Coulomb repulsion responsible for the correlation
phenomena.

\begin{figure}[b!]
\includegraphics[width=0.9\linewidth]{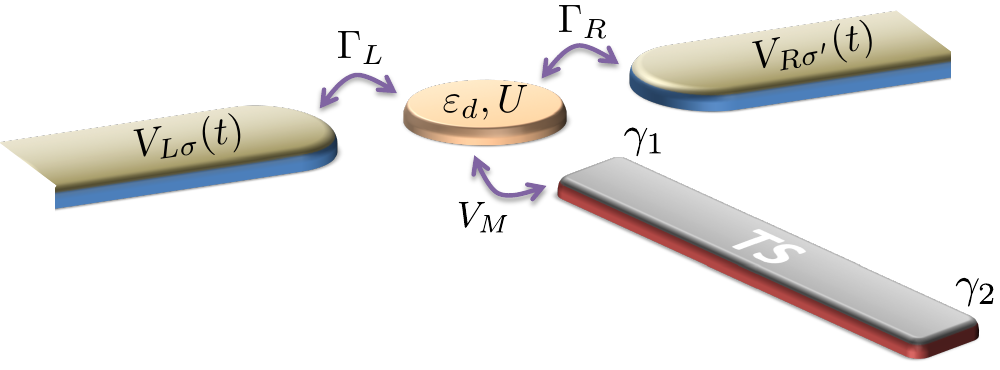}
\caption{Schematic of the hybrid setup with a correlated quantum dot 
    placed between the external (left and right) electrodes and 
    side-coupled to the topological superconducting (TS) nanowire,
	hosting the Majorana end-modes (described by $\hat\gamma_{1}$ and $\hat\gamma_{2}$).
	The quantum dot is characterized by the energy level $\e_d$ and  
	the Coulomb potential $U$. The dynamical charge transport is driven by
	time-dependent voltage $V_{j\s}(t)$ applied between the contacts.}
\label{fig:system}
\end{figure}	

In this paper we investigate dynamical hallmarks of the Majorana mode 
coupled to the strongly correlated quantum dot that would be observable in 
charging-discharging processes driven by the external {\em ac} field.
For microscopic considerations we choose the single dot hybrid structure, 
displayed in Fig.~\ref{fig:system}.
Here proposed methodology can be naturally extended to other, more complex structures.
As an example, we point out some alternative realization of our model (see Appendix \ref{sec:2QD}), 
where specific contributions of the dynamical conductance could be
examined without spin-resolved measurements.

Hybridization of the quantum dot with a topological superconducting 
wire gives rise to a leakage of the Majorana boundary mode onto 
the side-attached QD \cite{Vernek2014Apr}. When confronted with the 
Coulomb interactions it develops the unique \emph{Kondo-Majorana} 
(KM) strong coupling low-energy fixed point 
\cite{Cheng2014Sep,Silva2020Feb,Majek2022Feb}
characterized by universal spin-asymmetric spectral density \cite{Lee2013Jun},
resembling the one in non-interacting systems, yet revealing a fully-screened 
impurity magnetic moment \cite{Wojcik2023Mar}. The latter seems
difficult to be measured directly, whereas the static {\em dc}-transport 
features of such KM fixed point are hardly distinguishable from those 
of noninteracting Majorana-quantum dot hybrid systems \cite{Ptok2017Nov,Leumer2020Aug,Majek2022Oct,Diniz2023}.
Here we show that {\em ac} transport properties would be a suitable tool to discriminate them.

In what follows, we analyze the universal {\em ac} conductance over extended 
range of frequencies, bounded from above by the energy scale corresponding 
to the Majorana coupling, 
$\w_{\rm max } \sim 2\Gamma_{\M}$, 
and from below by the scale related to the overlap of Majorana modes, $\e_M$,
encountered in short-length topological superconducting systems, 
${\w_{\rm min } \sim \e_\M (1+\Gamma_\M/\e_\M)^{-1}}$. 
Here, $\Gamma_{\M} \approx 2 V_\M^2/\Gamma$, where $V_\M$
is the matrix element between the QD and TS, while $\Gamma$
denotes the coupling strength to the left and right leads.
In the case of long topological superconducting nanowire, $\e_\M \to 0$, 
the universal feature corresponds to the zero-bias peak,
and $\w_{\rm max}$ is the maximal 
{\em ac} frequency allowing for its observation.
Taking the Coulomb interaction effects into account,
we carefully investigate what dynamical response characterizes the Kondo-Majorana fixed point,
which (to the best of our knowledge) is still awaiting its experimental observation. 
With the present paper we aim to provide means of such detection.

The paper is organized as follows. 
In \Sec{model} we introduce the model and describe the numerical methods, 
based on numerical renormalization group (NRG) technique \cite{WilsonNRG}
and Kubo formalism \cite{Kubo1966Jan}.
Section~\ref{sec:GF} discusses the relevant energy scales, inferred from 
analysis of all contributions to the response functions.
Then, \Sec{G} combines these contributions into differential conductance 
of the device for various biasing scenarios. 
In \Sec{other} we briefly discuss alternative methods
capable to deal with correlation phenomena under {\em ac} nonequilibrium
conditions. Finally, \Sec{conclusions}
concludes our study.
Additionally, Appendix \ref{sec:2QD} presents an alternative experimental 
setup for verification of our findings and Appendix \ref{sec:SC}
shows the results obtained when the topological superconductor is 
replaced by a conventional one.

\section{Theoretical description}
\label{sec:model}

The considered nanostructure (Fig.~\ref{fig:system}) consists 
of a correlated quantum dot placed between two external leads
and additionally coupled to a topological superconducting nanowire,
hosting the Majorana boundary modes described by the operators
$\op\gamma_1$ and $\op\gamma_2$, respectively.
External leads are subject to a time-dependent bias voltage,
inducing {\em ac} charge transport via the quantum dot.
In what follows, we briefly specify the low-energy microscopic 
model of this system and present the theoretical framework 
for treating the dynamical phenomena within the linear-response 
theory.

\subsection{Hamiltonian}
\label{sec:H}

The system's Hamiltonian consists of the following terms
\begin{equation}
    \op{H}=\sum_{j={\rm L,R}} \left( \op{H}_{j}
    + \op{H}_{j{\rm -QD}} \right) + \hat{H}_{\rm QDM}.
\label{hamil_system}
\end{equation}
The external electrodes are assumed to be reservoirs of the itinerant electrons
\beq
\op{H}_{j} = \sum_{\s} \int [\e - \mu_{j\s}(t)]\; \ck_{j\e\s}\ca_{j\e\s} d\e,
\eeq
where $\ck_{j\e\s}$ ($\ca_{j\e\s}$) denote the creation (annihilation)
operators of spin-$\sigma$ electrons in $j$-th lead,
which satisfy anticommutation relation ${\{\ck_{j\e\s},\ca_{j'\e'\s'}\} = \delta_{jj'}\delta_{\s\s'}\delta(\e - \e')}$,
and $\mu_{j\sigma}(t)$ is the corresponding time-dependent chemical potential.
Electron tunneling between the $j$-th lead and the quantum dot is described by
\beq
\hat{H}_{j{\rm -QD}} = \sum_{\s} \int \sqrt{\rho_j(\e)}
		\left( v_{j} \dk_{\s}\ca_{j\e\s} + \hc \right)d\e ,
\label{HT}
\eeq
where $v_{j}$ is the hopping matrix element and $\rho_j(\e)$ 
is the density of states of $j$-th lead.
Since our considerations refer to a narrow energy region
(of a width $\sim$ meV) inside the topological gap of 
superconducting nanowire, we assume the density of states to be flat,
$\rho_j(\e)=\rho\equiv 1/2D$, where $D$ is the band energy cut-off,
used as a convenient energy unit throughout, $D\equiv 1$.
The hybridization effects (\ref{HT}) can be taken into account
by introducing the couplings $\Gamma_{j} = \pi |v_{j}|^{2} \rho_j$.

The last term of the Hamiltonian (\ref{hamil_system}) describes 
the quantum dot combined with the Majorana modes of the 
TS \cite{Leijnse2011Oct,Liu2011Nov}
\beqa
\label{HDM}
\op{H}_{\rm QDM} &=& \e_d \sum_{\sigma} \op{n}_{\sigma}
		+ U \op{n}_{\up}\op{n}_{\down} 
		\nonumber \\&+&	
		V_{\rm M} (\dk_\down - \da_\down )\op\gamma_1
		+ i \e_{\rm M} \op\gamma_1 \op\gamma_2,
\label{QD+Majorana}
\eeqa
where $\dk_{\s}$ ($\da_{\s}$) is the creation (annihilation)
operator for spin-$\sigma$ electrons on the QD of energy $\e_d$
and Coulomb correlations $U$,
$\op{n}_{\s} = \dk_{\s}\da_{\s}$ is the corresponding electron number operator,
and $\hat{\gamma}_{1,2}$ are the operators of the boundary zero-energy modes.
Hybridization of the quantum dot spin-$\down$ electrons 
with the left hand-side Majorana quasiparticle is denoted by $V_\M$,
while $\e_\M$ stands for an overlap between the Majorana modes.
It is useful to represent the Majorana operators $\hat{\gamma}_{i}$ in
terms of the conventional fermion operators $\hat{f}_{\rm M}^{(\dagger)}$,
\begin{subequations}
\beqa
\op\gamma_1 \es (\fa_{\rm M} + \fk_{\rm M})/\sqrt{2}, \\
\op\gamma_2 \es -i (\fa_{\rm M} - \fk_{\rm M})/\sqrt{2},
\eeqa
\end{subequations}
which obey the standard anti-commutation relations.

\subsection{AC conductance}
\label{sec:AC}

The charge current flowing from the $j$-th lead
to the quantum dot in spin channel $\s$ can be expressed by the operator
${\op{I}_{j\s}(t) = e \, \partial_t \op{N}_{j\s}(t)}$,
where ${\op{N}_{j\s} = \int \ck_{j\e\s}\ca_{j\e\s} \de\e }$
counts total number of spin-$\sigma$ electrons in the aforementioned lead,
time argument indicates the Heisenberg picture,
and $e = |e|$ stands for the elementary charge.
Assuming the topological superconductor to be grounded,
the applied time-dependent voltage affects the chemical potentials of external leads as
$V_{j\sigma}(t)\equiv -\mu_{j\sigma}(t)/e$.
Within the linear-response Kubo formalism \cite{Kubo1966Jan},
it is useful to define the Fourier transform
${V_{j\s}(t) = (2\pi)^{-1}\int e^{-i\w t} V_{j\s}(\w)\de \w}$. 
The expectation value of the charge current, $I_{j\s}(t) \equiv \avg{\op{I}_{j\s}(t)}$,
can be then expressed in terms of the Fourier transforms by \cite{Toth2007Oct,Moca2010Jun}
\beq
I_{j\s}(\w) = \sum_{j'\s'} \mathcal{G}_{jj'}^{\s\s'}(\w) V_{j'\s'}(\w),
\label{Iw}
\eeq
with the system's admittance
\beq
\mathcal{G}_{jj'}^{\s\s'}(\w) = \frac{i}{\w} 
	\left(  \GF{\op{I}_{j\s} |\op{I}_{j'\s'}}^{\rm ret}_\w - \GF{\op{I}_{j\s}| \op{I}_{j'\s'}}^{\rm ret}_{\w=0}  \right) ,
\label{admittance}
\eeq
where $\GF{\op{I}_{j\s} |\op{I}_{j'\s'}}^{\rm ret}_\w$ is the Fourier-transform
of the retarded Green's function of the current operator,
${\GF{\op{I}_{j\s} |\op{I}_{j'\s'}}^{\rm ret}_t = -i\Theta(t) \avg{[\op{I}_{j\s}(t) ,\op{I}_{j'\s'}(0)]}}$
(we set the time units such that $\hslash\equiv 1$ throughout the paper).
In what follows, we shall focus on the properties of the frequency-dependent conductance of the system
\begin{equation}
	G_{jj'}^{\s\s'}(\w) = \Re\;\left\{\mathcal{G}_{jj'}^{\s\s'}(\w)\right\} = 
	-\frac{1}{\w} \Im \;\left\{  \GF{\op{I}_{j\s} |\op{I}_{j'\s'}}^{\rm ret}_\w  \right\}.
\end{equation}
The current operator for our setup is given by
\beq
\op{I}_{j\s} = i e\left( v_{j} \op\psi^{\dagger}_{j\s} \da_\s - \hc \right) 
\eeq
with 
\beq
\op\psi_{j\s}^\dag = \int \! \sqrt{\rho_j} \; \ck_{j\e \s} \de\e 
\eeq
being the corresponding field operator for the creation of a spin-$\sigma$ electron in the $j$-th lead.
We introduce an effective tunnel-matrix element,
$v = \sqrt{v_{L}^2+v_{R}^2}$, and perform a transformation from the {\em left-right} to the {\em even-odd} basis
\beqa
\op\psi_\s^{\rm e} \es \frac{v_{L}}{v} \op\psi_{L\s} + \frac{v_{R}}{v} \op\psi_{R\s} , \\
\op\psi_\s^{\rm o} \es -\frac{v_{R}}{v} \op\psi_{L\s} + \frac{v_{L}}{v} \op\psi_{R\s} .
\eeqa
Then, for the even (odd) current operator one finds
\begin{equation}
	\op{I}_{\s}^{\rm e(o)} = iv e \;\op{\mathcal{I}}_{\s}^{\rm e(o)}, 
\end{equation}
with
\begin{equation}
  \op{\mathcal{I}}_{\s}^{\rm e(o)} = \op\psi^{{\rm e(o)}\dagger}_\s \da_\s  -  \dk_\s \op\psi^{{\rm e(o)}}_\s ,
\end{equation}
such that $\op{I}_{j\s}$ can be expressed as
\beq
\op{I}_{j\s} 
	= ie \left[
		\frac{v_{j}^2}{v}\op{\mathcal{I}}^{\rm e}_\s 
		+(-1)^{\delta_{j,L}}\frac{v_{L}v_{R}}{v}\op{\mathcal{I}}^{\rm o}_\s
		\right]. 
\eeq
Thanks to $\GF{\op{\mathcal{I}}^{\rm e}_\s | \op{\mathcal{I}}^{\rm o}_{\s'}} = 
\GF{\op{\mathcal{I}}^{\rm o}_\s | \op{\mathcal{I}}^{\rm o}_{-\s}} = 0$, one obtains
\beqa
\frac{G_{jj'}^{\s\s'}(\w)}{G_0} \es
-\delta_{\s\s'}\eta_{jj'}  \frac{2 \Gamma_{L}\Gamma_{R}}{\Gamma} 
\frac{1}{\w}
\Im \left[\GF{\op{\mathcal{I}}^{\rm o}_{\s} | \op{\mathcal{I}}^{\rm o\dagger}_{\s}}^{\rm ret}_\w\right]
\nn\\&&
- \frac{2\Gamma_{j}\Gamma_{j'}}{\Gamma} 
\frac{1}{\w}
\Im \left[\GF{\op{\mathcal{I}}^{\rm e}_{\s} | \op{\mathcal{I}}^{\rm e\dagger}_{\s'}}^{\rm ret}_\w\right] 
,\qquad
\label{Gjj_2}
\eeqa
where $\Gamma = \Gamma_L + \Gamma_R$, $G_0 = 2e^2/h$,
and $\eta_{jj'} = 1$ if $j=j'$ and $\eta_{jj'}=-1$ otherwise.

Note that $\op\psi_\s^{\rm o}$ is not present in the tunneling Hamiltonian, 
\eq{HT}, \ie{} it is just a free fermionic field whose Green's functions are known exactly.
Thus, one can relate the odd contribution to the quantum dot spectral function
through the Wick theorem without introducing any approximations.
On the other hand, the current correlation function associated with the even channel
needs to be determined explicitly. Then, the formula for the 
frequency-dependent conductance can be
written in a more compact form as
\cite{Toth2007Oct,Moca2010Jun,Plominska2017Apr}
\beq
\frac{G_{jj'}^{\s\s'}(\w)}{G_0} = \delta_{\s\s'}\eta_{jj'}  \frac{2\Gamma_{L}\Gamma_{R}}{\Gamma^2} g^{\rm o}_\s(\w)
+ \frac{2\Gamma_{j}\Gamma_{j'}}{\Gamma^2} g^{\rm e}_{\s\s'}(\w)
\label{Gjj}
\eeq
with the functions
\begin{equation}
g^{\rm o}_\s(\w) \!=\!\frac{\Gamma}{2\omega}
 \!\int\!\! \Im\; \GF{\da_\s | \dk_\s}^{\rm ret}_{\w'} 
[f(\w' \!+ \w) \!-\! f(\w' \!- \w)]\de\w',
\label{go}
\end{equation}
and
\begin{equation}
g^{\rm e}_{\s\s'}(\w) \!=\!
-\frac{\Gamma}{\w}\Im\; \GF{ \op{\mathcal{I}}^{\rm e}_{\s} | \op{\mathcal{I}}^{{\rm e \dagger}}_{\s'} }^{\rm ret}_\w
,\quad\;
\label{ge}
\end{equation}
where $f(\w)=\left[1+\exp{(\w/T)}\right]^{-1}$ is the Fermi-Dirac distribution function
($k_B\equiv 1$).

\subsection{Treatment of the correlations}
\label{sec:NRG}

To deal with the correlation effects in the low-temperature regime
where the Kondo effect emerges, we make use of the NRG approach
\cite{WilsonNRG,fnrg,Toth2008Dec} in its full-density-matrix implementation \cite{Anders2005},
which allows for sum-rule conserving determination of the spectral functions 
\cite{Weichselbaum} relevant for our study.
This method reliably yields the set of low and high energy eigen-states,
and the Green's functions can be calculated from
the Lehmann representation, imposing a broadening with the log-Gaussian kernel \cite{Weichselbaum}.
We also use the $z$-averaging trick \cite{Z} with $8$ different values of $z$,
which allows to avoid over-broadening of spectral functions. This is 
particularly important for correct calculation of the even contribution, where 
${\GF{ \mathcal{I}^{\rm e}_{\s} | \mathcal{I}^{{\rm e}}_{\s'} }^{\rm ret}_\w/\w}$
needs to be obtained at small $\w$, cf.~\eq{ge}.
For specific computations we have chosen the following model parameters:
the discretization parameter $\Lambda=2$,
the number of states kept per iteration $N_K=2048$,
and the broadening width in the range $0.2 < b < 0.35$.
For values of the physical parameters we take $U=0.1 D$ and $\Gamma=U/10$,
which allow for clear presentation of numerical results. 
These remain qualitatively valid as long as the 
strong coupling regime is successfully reached, 
and the relevant energy scales are outlined in the following
section.
We focus on the zero-temperature case ($T=0$).

\section{Response functions and relevant energy scales}
\label{sec:GF}

As can be seen from the above discussion, the  behavior
of the $ac$ conductance is essentially determined by two 
dimensionless conductances $g^{\rm o}_\s(\w)$ and $g^{\rm e}_{\s\s'}(\w)$.
Therefore, it is of great importance to analyze
those functions separately.
This will be crucial in understanding different contributions
to $G_{jj'}^{\s\s'}(\omega)$ as well as
\begin{equation}
	G_{jj'}(\omega) = \sum_{\sigma\sigma'} G_{jj'}^{\s\s'} (\omega)
\end{equation}
depending on a way how the system is biased.

\begin{figure}[t!]
	\includegraphics[width=0.9\columnwidth]{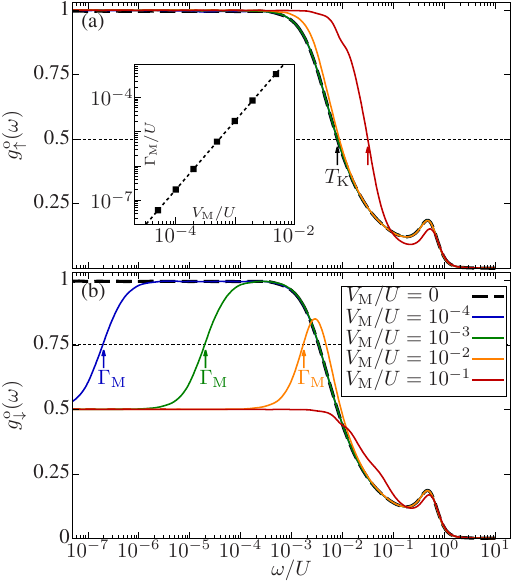}
	\caption{The odd contribution to the {\em ac} conductance 
		$g^{\rm o}_\s(\w)$ plotted as a function of frequency
		for different values of $\VM$, as indicated.
		(a) The spin-$\up$ component, with the $V_{\rm M}$-dependent 
		Kondo scale $T_{\rm K}$ indicated with arrows.
		(b) The spin-$\down$ component, with arrows indicating the relevant
		$\Gamma_{\rm M}$ scale (see the main text for details).
		Inset shows $\Gamma_{\rm M}$ as a function of $V_{\rm M}$,
		with dashed line corresponding to \eq{GammaM}.
		The parameters are: $U=0.1D$, $\e_d = -U/2$, $\Gamma = U/10$, and 
		$\e_{\rm M}=0$.
	}
	\label{fig:godd}
\end{figure}

\subsection{Odd response function}
\label{sec:go}

We start the discussion from the {\em odd} dimensionless 
conductance, cf.~\eq{go}, presented in \fig{godd}.
It describes the contribution from the processes that change
sign upon the left-right leads exchange, thus corresponding to 
the transport through the nanostructure from L to R lead.
These processes are governed mainly by the local spectral 
density of QD, symmetrized through the convolution
with the corresponding Fermi functions, cf.~\eq{go}.
Note that the spin symmetry is broken only by the coupling 
to spin-polarized TS quantified by $\VM$, which determines 
the spin quantization axis.
Therefore, $g^{\rm o}_\up(\w)$ exhibits a hump at $\w\sim U$, 
a minimum corresponding to Coulomb blockade for $\w \lesssim U$
and a peak for frequencies below the Kondo temperature $\TK$, here defined through the half-width 
\beq
g^{\rm o}_\up(\w=\TK) = 0.5.
\label{TKdef}
\eeq
As long as $\TK > \VM$, it is practically independent of $\VM$,
and therefore the well-known estimation \cite{Haldane_Phys.Rev.Lett.40/1978} 
remains valid,
\beq
\TK^0 = \sqrt{\frac{U\Gamma}{2}}\exp\left[ \frac{\pi}{2}\frac{\e_d(\e_d+U)}{U\Gamma} \right].
\label{TK0}
\eeq
For larger $\VM$, however, the results indicate increase of 
the Kondo scale. Similar tendency has already been reported in earlier 
numerical studies based on the spectral densities and temperature dependence of 
the {\em dc} conductance \cite{Ruiz-Tijerina2015Mar,Weymann2017Apr,Wojcik2023Mar}, 
but this goes against intuition concerning a competition between Kondo and Majorana couplings as well as
expectations of lack of such dependence from approximate RG schemes
\cite{Cheng2014Sep,Silva2020Feb,Baranski2023Jul}.
The issue has therefore remained controversial and is fully 
resolved in favor of $\TK$ increase only through the analysis of 
the even contribution to the {\em ac} conductance,
see the discussion of \eq{TK} in the sequel.

The other energy scale known from the analysis of conventional 
spectral functions of Majorana devices is the effective 
coupling strength to TS \cite{Lee2013Jun}
\beq
\GM \approx 2V_{\rm M}^2 / \Gamma .
\label{GammaM}
\eeq
It determines the energy scale below which the 
typical Majorana spectral signatures are visible.
Namely, at $\w \ll \GM$ the conductance in spin channel 
coupled to TS [$\s=\down$, cf.~\fig{godd}(b)] is pinned to $e^2/2h$. 
When $\GM < \TK$, the value of $\GM$ can be recognized from 
the condition $g^{\rm o}_\down(\w=\GM)=0.75$ \cite{Weymann2020Jan}. 
Then, the numerical calculations give results in agreement with \eq{GammaM}, 
cf.~the inset in \fig{godd}(a). 

\subsection{Even response function}
\label{sec:ge}

\begin{figure}[t!]
	\includegraphics[width=0.9\columnwidth]{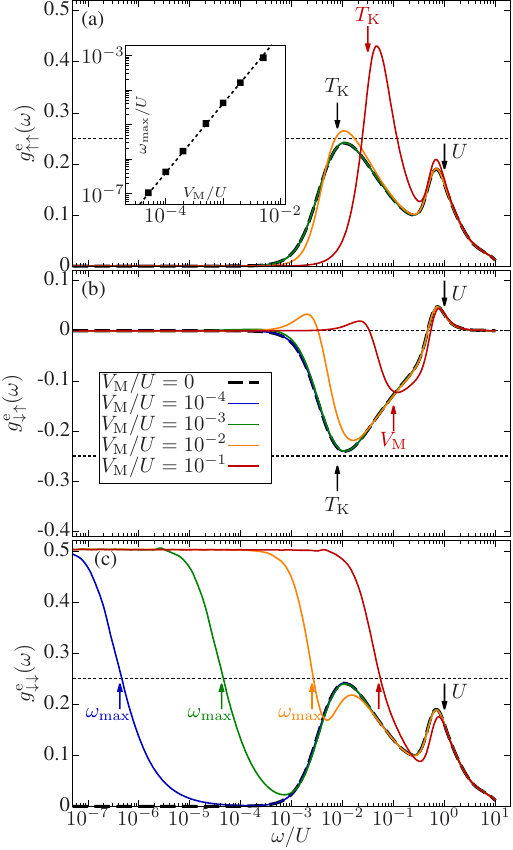}
	\caption{The even contribution to the {\em ac} conductance 
		$g^{\rm e}_{\s\s'}(\w)$ plotted as a function of frequency. 
		(a) presents $g^{\rm e}_{\up\up}(\w)$,
		(b) shows $g^{\rm e}_{\down\up}(\w) = g^{\rm e}_{\up\down}(\w)$,
		while (c) displays $g^{\rm e}_{\down\down}(\w)$
		calculated for different values of $\VM$, as indicated.
		The arrows indicate the relevant energy scales.
		Inset shows $\omega_{\rm max}$ as a function of $V_{\rm M}$,
		with dashed line corresponding to $\w_{\rm max}=2\GM = 4 \VM^2/\Gamma$.
		The parameters are the same as in Fig.~\ref{fig:godd}.
	}
	\label{fig:geven}
\end{figure}

While $g^{\rm o}_\s(\w)$ provides comparable insight into 
the system's properties as the local quantum dot spectral density, the
{\em even} contribution, $g^{\rm e}_{\s\s'}(\w)$, probes a
different Green's function, cf.~\eq{ge}, and complements the
information available from the local spectroscopy. This contribution
is even in the left-right leads exchange, so it corresponds to 
the processes of charging and discharging the quantum dot,
and transport from normal leads to the topological superconductor.
The two spin indices of $g^{\rm e}$ correspond to the response
and voltage bias, and the cross-terms appear as a consequence 
of effective spin exchange interactions induced by Coulomb 
correlations in the Hamiltonian, \eq{QD+Majorana}.
Even though in practice a sum over spins is usually measured,
we find it useful to discuss each component separately first.
We note that such components could be measured
in a double quantum dot setup without the need to resorting
to spin-dependent measurements, see Appendix \ref{sec:2QD}.

\begin{figure}[tb!]
	\includegraphics[width=0.9\linewidth]{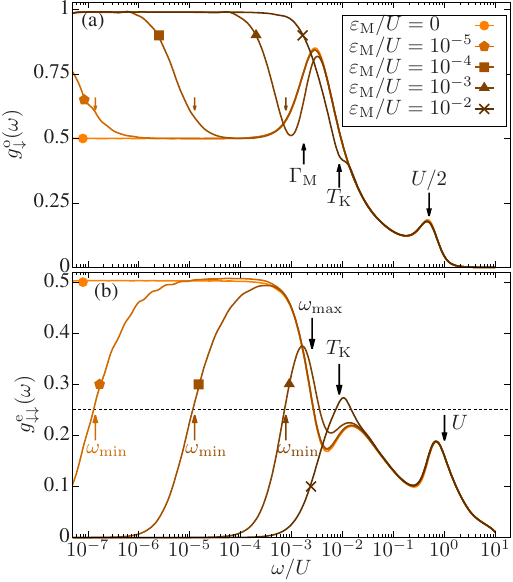}
	\caption{(a) The odd and (b) even contributions to the spin-$\down$ 
		response function due to a spin-$\down$ voltage calculated for 
		$\VM=10^{-2}U$ and a few chosen values of $\e_{\rm M}$,
		as indicated. The other parameters the same as in \fig{godd}.
		The arrows indicate relevant energy scales, with unlabeled
		arrows in (a) corresponding to relevant $\w_{\rm min}$
		shown in (b).
	}
	\label{fig:fig4}
\end{figure}

As can be seen in \fig{geven}(a), $g^{\rm e}_{\up\up}(\w)$ exhibits
a signal characteristic of the Kondo effect, \ie{} a peak at $\w\sim\TK$, 
and a second peak for $\w\sim U$ \cite{Moca2010Jun}.
Note that these features do not occur for a resonant model, therefore
they are unique signatures of Coulomb interactions and the resulting Kondo effect.
Similarly as in \fig{godd}(a), one observes now an increase of
$\TK$ with $\VM$. This enhancement of $\TK$ persists even 
if the wire is not long enough to make the Majorana overlap $\e_{\rm M}$ negligible,
see \fig{fig4}, and is most clearly shown in \fig{fig5}(a).
For different values of $\VM$, from our numerical data we find approximately 
\beq
\TK \approx \sqrt{ \left( \TK^0\right)^2 + 0.1 \VM^2},
\label{TK}
\eeq
which remains valid when the TS wire is short, $\e_{\rm M}\neq 0$, 
cf.~\fig{fig4}(a) and \fig{fig5}(a).

In \fig{geven}(b) the opposite-spin-response is shown 
[note that $g^{\rm e}_{\up\down}(\w)=g^{\rm e}_{\down\up}(\w)$ 
follows from the definition of relevant 
Green's functions, cf.~\eq{ge},
and the bosonic character of current operators]. 
Already for $V_M=0$, it possesses a set of unique features \cite{Moca2011Dec}.
It exhibits a positive peak for $\w\sim U$ 
and a negative one at $\w\sim\TK$.
This negative sign means that for the corresponding frequencies
the voltage in spin channel $\s$ induces a negative current
in the other spin channel, which could be 
seen as a consequence of the Kondo singlet oscillations.
When a spin-$\up$ electron leaves the quantum dot,
its singlet companion secures single-occupancy of the dot.
This is further confirmed by the fact that for $\VM=0$
the negative peak of $g^{\rm e}_{\down\up}(\w)$
is a mirror-image of the corresponding positive peak in $g^{\rm e}_{\s\s}(\w)$,
visible in Figs.~\ref{fig:geven}(a) and (c); see also the discussion of \fig{Geven}.
This means that the charge current in one spin channel is compensated by 
the opposite current of the other spin channel, which is
the essence of an antiferromagnetic spin exchange. 
When $\w\sim\TK$, the spin exchange is almost in resonance with 
the driving frequency, leading to 
fractional value of the components of the even conductance,
$\eta_{\s\s'} g^{\rm e}_{\s\s'} = 1/4$
and the unitary value of the spin conductance.\footnote{%
To see that this is indeed the value relevant in the unitary regime, 
let us assume that the corresponding
$\w$-dependent spin bias $V_{j\up}(\w)=-V_{j\down}(\w)=V^{\rm e}(\w)/2$
is applied to both normal leads. Then, the charge currents fulfill,
$I_{j\s}(\w) = (G_0/2) V^{\rm e}(\w) [g^{\rm e}_{\s\up}(\w)-g^{\rm e}_{\s\down}(\w)]$,
and the total spin current from normal electrodes is
$I_\mathcal{S}(\w) = \sum_{j\s} \s I_{j\s}(\w) = G_0 V^{\rm e}(\w) [	 g^{\rm e}_{\up\up}(\w)		-g^{\rm e}_{\up\down}(\w)	
											-g^{\rm e}_{\down\up}(\w)	+g^{\rm e}_{\down\down}(\w)]$.
Consequently, $I_\mathcal{S}(\w)$ becomes $G_0 V^{\rm e}$ at resonance,
i.e. for $\w\approx\TK$, which is indeed the unitary value.}
When $\w$ significantly exceeds $\TK$, the process becomes
inefficient, as spin exchange no longer keeps up to rapidly
oscillating driving bias.

For $0 < \VM \leq 10^{-2}U$, the picture changes mainly 
quantitatively: the negative peak position shifts toward 
higher $\w$ as a consequence of increase of $T_K$.
However, it should be noted that for $\VM \gtrsim \TK$, while the 
peak in $g^{\rm e}_{\up\up}(\w)$ becomes higher in the presence 
of TS, the negative peak in $g^{\rm e}_{\down\up}(\w)$ is reduced 
and supplemented with a small positive one at lower $\w$; 
cf. the curves for $\VM=10^{-2}$ in \fig{geven}(a-b). This 
could be understood as the competition between the Kondo exchange
and Andreev processes, the latter ones relevant here only 
for the spin-down electrons. It becomes even more apparent for 
larger $\VM$, $\VM \gtrsim \TK^0$, when the positions of the peaks
become different:
in $g^{\rm e}_{\up\up}(\w)$ the peak remains at $\TK(\VM)$, while in 
$g^{\rm e}_{\down\up}(\w)$ it follows simply $V_M$ [cf.~the curve 
for $\VM=10^{-1}U$ in \fig{geven}(a)],
which is the energy scale of the QD-TS hopping. This leads to 
$g^{\rm e}_{\up\up}(\w\approx\VM) < -g^{\rm e}_{\up\down}(\w\approx\VM)$,
and thus one obtains a {\em negative} total response in the spin-down
channel, cf.~\fig{Geven}.

The most prominent result for the even response function is presented in \fig{geven}(c),
which shows the spin-$\down$ response to the spin-$\down$ bias,
i.e. $g^{\rm e}_{\down\down}(\w)$.
Besides the Kondo-related peak identical to the one visible in $g^{\rm e}_{\up\up}(\w)$,
$g^{\rm e}_{\down\down}(\w)$ acquires a non-zero value in the static limit ($\w\to 0$).
This means steady current flowing into QD at a constant bias and can be understood
as the current flowing into TS. This result extends to higher $\w$,
in fact to $\w\sim \w_{\rm max} = 2\GM \approx 4\VM^2/\Gamma$, as indicated in \fig{geven}(c)
with upward arrows. The magnitude of $\w_{\rm max}$ is further analyzed in \fig{fig5}(b).

\subsection{Role of Majorana overlap}
\label{sec:epsM}

\begin{figure}[tb!]
\includegraphics[width=0.9\linewidth]{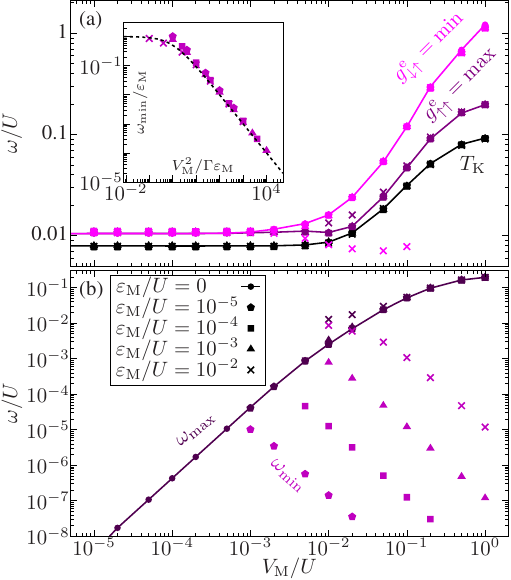}
\caption{(a) The Kondo scale $\TK$, defined in \eq{TKdef}, 
		     estimation of the Kondo scale 
		     determined as the position of the maximum in $g^{\rm e}_{\up\up}(\w)$,
		     and the position of the minimum in $g^{\rm e}_{\down\up}(\w)$,
		     as functions of $\VM$. 
		 (b) $\w_{\rm max}$ and $\w_{\rm min}$ as functions of $\VM$.
		 In both panels, lines correspond to $\e_{\rm M}=0$, while different 
		 point styles correspond to different values of $\e_{\rm M}$, as indicated.
		 Inset in (a) shows $\w_{\rm min}$ for all considered $\e_{\rm M}$,
		 appropriately scaled, such that they all collapse to a single curve
		 estimated by \eq{wMin} and indicated with a dashed line.
		 }
\label{fig:fig5}
\end{figure}

The topological superconducting nanowires should be sufficiently long
in order to prevent any overlap between the Majorana zero-energy modes.
In practice, however, this may be difficult to achieve and a non-negligible
$\e_{\rm M}$ may exist, as assumed in \eq{QD+Majorana}.
This quite drastically changes the situation, because the Kondo-Majorana
fixed point is not stable against such perturbation, \ie{} $\e_{\rm M}$ is 
a \emph{relevant} perturbation in the RG sense.
Still, the latter determines the physics at the intermediate temperatures $T$,
namely when $|\e_{\rm M}| < T < \TK$ \cite{Majek2022Feb}.
This looks similar for the {\em ac} response, where $\w$ plays 
the role of $T$, as is shown in \fig{fig4}. In both $g^{\rm o}_{\down}(\w)$ and
$g^{\rm e}_{\down\down}(\w)$ there appears a scale,
denoted here as $\w_{\rm min}$,  whose magnitude can be estimated from
\beq
\w_{\rm min} =  \e_\M (1+\Gamma_\M/\e_\M)^{-1}.
\label{wMin}
\eeq
This is simply the energy scale related to crossing to a stable
fixed point, which for $\e_\M \neq 0$ is not the Majorana-Kondo
but rather the conventional Kondo fixed point.

As a summary of this section, all the relevant energy scales are plotted in \fig{fig5}.
In panel (a) it is shown that $\e_{\rm M}$ does not really influence 
$\TK$ and only for $\e_{\rm M}\sim\VM$ it may affect the position of $g^{\rm e}_{\down\up}(\w)$ minimum.
On the other hand, panel (b) presents how the range of Majorana regime in the {\em ac} response,
in particular $\w_{\rm min}(V_\M,\e_\M)$, changes with $\VM$ for different 
$\e_{\rm M}$.
The universality of the formula (\ref{wMin}) is demonstrated in the inset of \fig{fig5}(a). 

\section{{{\em AC} Conductance in different bias configurations}}
\label{sec:G}

Having discussed the odd and even response functions,
let us now examine the characteristics of the frequency-dependent conductance for three different bias configurations. 
In the first one, referred to as antisymmetric bias configuration,
the voltage is applied as, $-V_{{\rm L}\s}(\w)=V_{{\rm R}\s}(\w)\equiv V^{\rm o}(\w)/2$.
Then, assuming symmetrical coupling of the quantum dot to the leads,
$\Gamma_{\rm L}=\Gamma_{\rm R}$, we have 
\beq
-I_{{\rm L}\s}(\w) = I_{{\rm R}\s}(\w) = G^{\rm o}_\s(\w)V^{\rm o}(\w),
\eeq 
with the {\em ac} conductance given only by the odd response function
\beq
G^{\rm o}_\s(\w) = \frac{G_0}{2} g^{\rm o}_\s(\w).
\label{Go}
\eeq
This is in contrast to the symmetric bias configuration,
$V_{{\rm L}\s}(\w)=V_{{\rm R}\s}(\w)\equiv V^{\rm e}(\w)$
in which one can probe the even contribution to the frequency-dependent conductance.
In this case, the total {\em ac} current entering the quantum dot from
the normal leads can be written as
\beq
I_{{\rm L}\s}(\w) + I_{{\rm R}\s}(\w) = G^{\rm e}_{\s}(\w)V^{\rm e}(\w),
\eeq
with the conductance in spin $\s$ channel 
\beq
G^{\rm e}_{\s}(\w) 
= \sum_{jj'\s'}G_{jj'}^{\s\s'}
= 2G_0 	\left[ g_{\s\up}^{\rm e}(\w) + g_{\s\down}^{\rm e}(\w) \right] .
\label{Ge}
\eeq
In general, for $\Gamma_{\rm L}=\Gamma_{\rm R}$, any spin-independent 
voltage bias $V_{j\up}(\w)=V_{j\down}(\w)=V_j(\w)$
can be decomposed into the even and odd parts.
Still, we find it useful to discuss also the case
when the time-dependent voltage is applied to one lead, i.e. $V_{\rm R}=0$.
One then has
\beq
I_{j\s}(\w) = G^{jL}_{\s}(\w)V_{\rm L}(\w) 
\eeq
with 
\begin{eqnarray}
G^{\rm LL}_\s(\w) &=& \frac{G_0}{2} \left[ g_{\s\up}^{\rm e}(\w) + g_{\s\down}^{\rm e}(\w) + g^{\rm o}_\s(\w) \right] ,
\label{GLL}\\
G^{\rm RL}_\s(\w) &=& -\frac{G_0}{2} \left[ g_{\s\up}^{\rm e}(\w) + g_{\s\down}^{\rm e}(\w) - g^{\rm o}_\s(\w) \right] 
\label{GRL}
\end{eqnarray}
and the minus sign in \eq{GRL} follows from 
taking the current from the quantum dot to the right lead as positive.

In the static limit, $\w\to 0$, the net current flowing from the normal leads 
to the dot must drain into the superconducting nanowire.
In contrast, for $\w\neq 0$, 
the processes of charging and discharging can yield non-zero signal.
Note that we assume here the superconductor to be always grounded,
\ie{} no bias is ever applied to it.

In the following, we discuss the total conductances
$G^x(\w) = G^x_\up(\w)+G^x_\down(\w)$
and the corresponding spin polarization, given by 
\beq
\Pol^x(\w) 	
		 	= \frac{|G^x_\up(\w)|-|G^x_\down(\w)|}{|G^x_\up(\w)| + |G^x_\down(\w)|},
\label{Pol}
\eeq
where $x$ is one of the labels $\{{\rm e,o,LL,LR}\}$, instead 
of addressing each spin component separately.
It should be noted that in our model we do not include any 
magnetic field acting on the QD, so only the hopping to TS 
determines the spin quantization axis. This might be relevant
for TS nanowires achieved through deposition of magnetic adatoms on
the ferromagnetic substrate.
It may be also relevant for more conventional 
semiconductor-based wires, assuming the correspondingly different $g$--factor for the quantum dot.

\begin{figure}[tb!]
\includegraphics[width=0.9\linewidth]{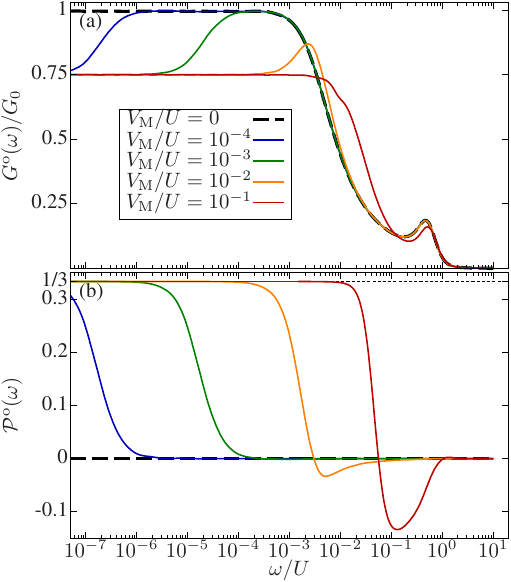}
\caption{(a) The total conductance between the normal leads and 
		 (b) its spin polarization calculated as function of $\w$
		 assuming small spin-independent bias applied antisymmetricaly to
		 the left and right leads.
		 The parameters are the same as in \fig{godd}.}
\label{fig:Godd}
\end{figure}

\subsection{The case of antisymmetric biasing}
\label{sec:Go}

Let us start with the situation when
an antisymmetric bias is applied to the system, \ie{} 
$-V_{{\rm L}\s}(\w) = V_{{\rm R}\s}(\w) = V^{\rm o}(\w)/2$.
In such a case, according to \eq{Go}, each spin component 
of the conductance is basically given by $g_\s^{\rm o}(\w)$, 
such that $G(\w)\equiv G^{\rm o}(\w)=\sum_\s G^{\rm o}_\s(\w)$.
While the behavior of $g_\s^{\rm o}(\w)$ has been discussed in the context of \fig{godd},
here we focus on the total {\em ac} conductance and its spin polarization,
which are presented in \fig{Godd}.
First of all, one can see that the zero-frequency total conductance
$G^{\rm o}(\w=0) = 3e^2/2h = (3/4)G_0$ for $\VM \neq 0$, 
cf.~\fig{Godd}(a). The order of magnitude of $\w_{\rm max}$ 
can be roughly estimated as the energy scale where this value is reached.

Any non-zero $\VM$ leads to spin imbalance, also at $\w=0$, 
where $g^{\rm o}_\up(0)=2g^{\rm o}_\down(0)$.
This leads to $\Pol^{\rm o}(\w\ll\w_{\rm max}) = 1/3$, 
as is visible in \fig{Godd}(b).
On the other hand, whenever $\VM \gtrsim \TK$, at $\w \sim \VM$, the 
spin-$\down$ conductance is enhanced in comparison to the spin-$\up$ 
one, due to additional TS-QD processes possible in this spin 
channel. Consequently, $\Pol$ becomes negative in this range of frequencies.

\begin{figure}[tb!]
\includegraphics[width=0.9\linewidth]{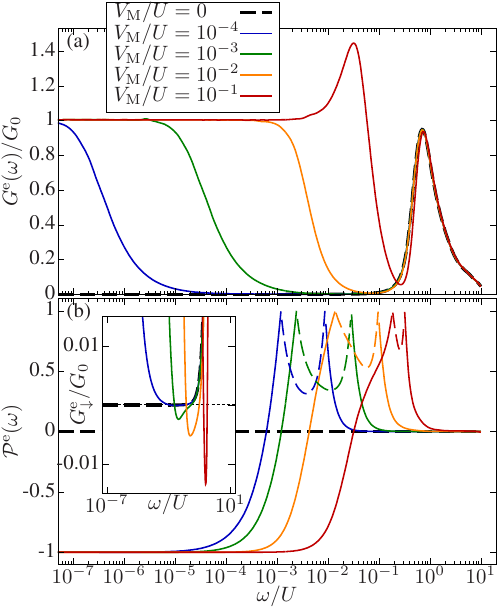}
\caption{(a) Total conductance between the leads and quantum dot
	and (b) the corresponding spin polarization calculated as function of $\w$.
	The dashed lines in (b) were used to indicate the range of 
	$\w$ where $G^{\rm e}_\down(\w)<0$.
    The inset shows $G^{\rm e}_\down(\w)$ in this regime.
    The parameters are as in \fig{geven}, with identical small 
	spin-independent bias applied to both left and right leads,
    $V_{{\rm L}\s}(\w) = V_{{\rm R}\s}(\w) = V^{\rm e}(\w)$.	
	}
\label{fig:Geven}
\end{figure}

\subsection{The case of symmetric biasing}
\label{sec:Ge}

In turn, we focus on the case of symmetric biasing, that is, we assume
$V_{{\rm L}\s}(\w) = V_{{\rm R}\s}(\w) = V^{\rm e}(\w)$
and define the conductance through the total current entering the quantum dot
from both normal leads. Note that in this biasing situation,
the total conductance is given by the even contribution only,
i.e. $G(\w)\equiv G^{\rm e}(\w)=\sum_\s G^{\rm e}_\s(\w)$.
The corresponding conductance is displayed in \fig{Geven}(a).
First of all, it exhibits a large peak at frequencies of the order 
of Coulomb interaction strength $U$, irrespective of $\VM$. This
signal is a signature of charging-discharging dynamics, and remains
the only feature relevant for $\VM=0$ case, \ie{} when QD is 
decoupled from TS. 
For $\VM>0$, however, the most prominent feature is a low-frequency 
($\w\lesssim \w_{\rm max}$) plateau of perfectly spin-polarized conductance
$2e^2/h$, cf.~the latter plotted in \fig{Geven}(b).
This is a signature of resonant Andreev transport through QD into 
the TS \cite{Law2009Dec,Flensberg2010Nov,Lutchyn2018May}, 
persisting at finite frequencies. Ultimately, for $\VM \gtrsim \TK$ 
(see the curve for $\VM=10^{-1}U$ in the figure) 
an additional peak at $\w\sim \TK$ is present. This is a consequence of 
separation between positions of positive peak in $g^{\rm e}_{\up\up}(\w)$,
present always at $\w\sim\TK$, and negative peak in $g^{\rm e}_{\up\down}(\w)$,
shifting from $\w\sim\TK$ for $\VM \lesssim  \TK$ toward $\w\sim\VM$ for $\VM \gtrsim \TK$;
see also the discussion of \fig{geven}. This additional peak increases
the total dynamical conductance over $G_0$ threshold and partially lifts
the spin polarization around $\w\sim\TK$. 

As has been pointed out in the context of \fig{geven}, already 
for $0 < \VM \lesssim T_K$, the compensation between $\w\sim\TK$ peaks
of $g^{\rm e}_{\up\up}(\w)$ and $g^{\rm e}_{\up\down}(\w)$ becomes
imperfect. While this is barely visible in the conductance curves,
which show almost completely suppressed transport, the plot of
spin polarization, \fig{Geven}(b), reveals vast domination of 
the spin-$\up$ channel transport. At two frequencies, the 
spin-$\down$ contribution vanishes, and $\Pol^{\rm e}(\w)=1$.
Between these frequencies $G^{\rm e}_{\down}(\w)$ even becomes 
genuinely negative, \ie{} the spin-$\down$ current flows in opposite 
direction than the spin-$\up$ one. This means a conversion of 
a conventional bias to the spin current, which might be interesting
from the point of view of applications.

\begin{figure*}[tb!]
	\includegraphics[width=0.65\linewidth]{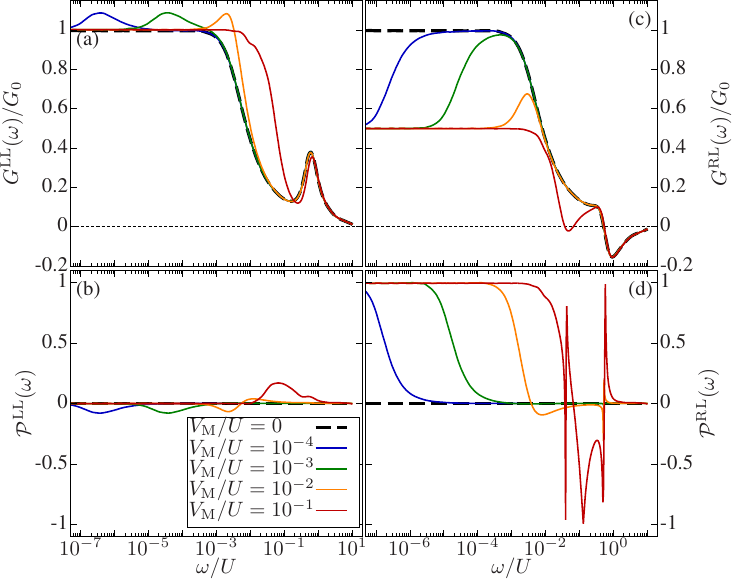}
	\caption{(a) The dynamical conductance from the left lead to the quantum dot
		and (b) the corresponding spin polarization, 
		as well as (c) the conductance from the right lead to the quantum dot
		together with (d) its spin polarization.
		The parameters are the same as in \fig{godd},
		with a time-dependent voltage applied to the left lead only.}
	\label{fig:GjL}
\end{figure*}

\subsection{The case of one-lead pumping}
\label{sec:GjL}

Finally, as an example of a general situation, where both odd and even 
channels are present, we consider {\em ac} bias applied to 
only one (left) lead. The conductance between this lead and the quantum dot
is given by \eq{GLL} and presented in \fig{GjL}(a) for selected values of $\VM$.
Clearly, it interpolates between $G^{\rm o}(\w)$ and $G^{\rm e}(\w)$, cf.~\fig{Godd}
and \fig{Geven}. It exhibits a peak at $\w\sim U$ and grows 
to approximately $2e^2/h$ for $\w\lesssim \TK$. Quite unfortunately,
most of the Majorana features get washed away, as a consequence
of the compensation between $G^{\rm e}(\w)$ increase and $G^{\rm o}(\w)$ 
suppression with increasing $\VM$. The only remaining features are small 
bumps appearing at frequencies $\w \sim \w_{\rm max}$. 
Similar is the fate of the spin polarization features, presented
in \fig{GjL}(b), where only a relatively low signal persists in this range.

Much more interesting features,
giving somewhat more insight into the system behavior,
are revealed in the response in the other lead,
which is proportional to $G^{\rm RL}(\w)$, cf. \eq{GRL}.
The corresponding conductance is presented in \fig{GjL}(c).
Now, one can see that at $\w\lesssim\w_{\rm max}$ the 
system works as an excellent spin filter. The electrons with both 
spins are pumped between the left lead and the dot.
On the other hand, topological superconductor allows only 
the spin-$\down$ channel to enter, meanwhile the unbiased 
right lead constitutes a drain for spin-$\up$ channel. 
At higher $\w$, $G^{\rm RL}(\w)$ drops below $0$.
This behavior is easiest to understand for $\w\sim U$,
when the processes of charging-discharging the quantum dot
are the most effective means of transport, and while most
of the charge is supplied by the biased lead, some part is 
also drawn from the other normal lead. Additionally, when
the peaks in $g^{\rm e}_{\up\up}(\w)$ and $g^{\rm e}_{\up\down}(\w)$ 
are separated (see the curves corresponding to $\VM = 10^{-1}U$ 
in \fig{GjL}), a complex interplay between all different 
contributions gives rise to an additional dip of the conductance 
visible in \fig{GjL}(c) and violent changes of $\Pol^{\rm RL}(\w)$
around the zeros of conductances in both spin channels, see \fig{GjL}(d).
Such sharp frequency-driven changes of the sign of spin 
polarization might be very useful for applications.

\section{Discussion}
\label{sec:other}

In this section, we point out possible extensions and limitations of our considerations, discuss other treatments using effective Hamiltonians and comment on the influence of disorder on Majorana quasiparticle features.

Correlation effects driven by the Coulomb repulsion between opposite-spin
electrons on an Anderson-type impurity can be described
under certain conditions within effective spin models, 
determining the exchange coupling perturbatively by canonical 
transformation \cite{Schrieffer_1966}. This approach proved to
be successful when applied to nanostructures with a correlated 
quantum dot placed between metallic lead and conventional superconductor, 
clarifying the subtle relationship of the proximity-induced 
electron pairing and the Kondo physics \cite{Domanski-2016}. 
Effective spin interactions of the Anderson impurity side-attached to the 
topological superconducting nanowire have been also investigated in 
this perturbative framework \cite{Cheng2014Sep,Silva2020Feb,Baranski2023Jul}.
Such study revealed that the Majorana mode affects the spin-exchange 
potential and additionally introduces the Zeeman field, lifting 
the spin degeneracy of quantum dot energy level.
In consequence, the spin-resolved Kondo effect of the quantum dot is profoundly altered  \cite{Lee2013Jun}. 
However, perturbative elimination of charge fluctuations at QD 
can be applied only when all Hamiltonian terms changing QD occupation 
[\ie{} $\Gamma$ and $V_M$, cf.~\eq{HT} and \eq{QD+Majorana}]
are indeed irrelevant, while the exchange interactions scale 
toward strong coupling fixed point. 
Therefore, without RG analysis including
charge fluctuation terms, validity of such effective low-energy
spin model is limited.
Here, we have avoided this difficulty simply by
taking charge fluctuations into account, which does not 
constitute any significant complication within the proposed 
computation scheme.

A possible route for treating the nonequilibrium effects induced
by a periodically varying voltage applied across the quantum dot 
would be the time-dependent Schrieffer-Wolff transformation. 
Such approach adopted to the correlated quantum 
impurity placed between two normal leads displayed the two-channel 
Kondo physics \cite{Eckstein_2017}, where impurity is screened by 
separate conduction bands, corresponding to parity-even and odd 
superpositions of the external leads. Upon varying the amplitude 
and frequency of a drive, the system can be tuned to the critical 
point with symmetric coupling of the impurity to both channels.
Approaching this critical point, the spin susceptibility increases 
logarithmically with time. In the regime where energy absorption is 
low, the time-evolution of the impurity spin indeed revealed
dynamics typical for the two-channel model \cite{Eckstein_2017}. 
This treatment is perturbative, so its validity beyond weak
coupling regime is not clear.

More generally, joint influence of the strong correlations and periodic driving can be investigated from the perspective of the Floquet theory, which can be regarded as time-equivalent of the spatial Bloch treatment. 
Using such an approach to the setup with the Kondo impurity embedded 
between two metallic leads, a coherent dressing 
of the driving field manifested by side replicas of the Kondo 
resonance of the averaged conductance has been predicted \cite{Bruch_2022}. Main virtue of this method is its applicability from the weak to strong driving and ability to deal with short voltage pulses.
The Floquet-Kondo engineering enables also derivation of the effective 
models with multichannel degenerate points (even though the starting 
Hamiltonian is a single channel one) \cite{Flint_2023}. The  emergent 
channels of various physical situations in presence of {\em ac} 
external fields can be controlled by changing e.g.\ polarization, 
frequency, and/or amplitude. The resulting multichannel Kondo models 
could host a plethora of exotic phenomena, including non-Abelian 
anyons \cite{Flint_2023} and novel types of nonequilibrium 
superconductivity \cite{Eckstein_2018}.
The Floquet-formalism picture of the Kondo-Majorana interplay 
would be a significant extension of the results discussed here, 
in principle lifting the characteristic Kubo formalism assumption
of weak driving. However, it would require equally significant 
development of the methodology to address such a complex, 
periodically driven, strongly correlated system without introducing uncontrolled (\eg{} perturbative) approximations.

The boundary modes harbored in topological phases should be immune to disorder, although in particular setups this situation can vary. 
For instance, disorder might play an important role by inducing 
local features resembling the topological boundary states. For this reason, additional checks, \eg{} by inspecting the behavior of the {\em ac}-conductance, could resolve the true nature of the zero-energy quasiparticles. In particular, robustness of the Majorana modes of the inhomogeneous Rashba nanowire deposited on superconducting substrate has been discussed in Ref.\ \cite{Maska_2017} (see also other papers cited therein). 
It seems, however, that magnetic field (magnetic moments) 
would be useful because it induces a splitting of the trivial 
bound states, while the Majorana mode stays robust at zero energy. Thus, in a regime of the magnetic field corresponding to the topologically nontrivial superconductivity, there is some possibility to discriminate between the Majorana and disorder-induced quasiparticles.

\section{Conclusions and outlook}
\label{sec:conclusions}

We have investigated the dynamical charge transport 
through the correlated quantum dot side-attached to 
the topological superconducting nanowire, focusing 
on the strong coupling regime.
We have identified the frequency range where the signatures 
of Majorana modes occupying the ends of the wire can be observed. 
Our analysis of all the contributions to conductance
between the normal leads and the quantum dot revealed
that in general total conductance is a combination
of two contributions: the odd and even in
exchanging the left and right leads, respectively.
We have proposed several biasing schemes allowing for
addressing each of these contributions
separately, and we also discussed the case of driving 
only one of the normal leads when both contributions are relevant.

We have shown that at low frequencies the even 
contribution remains non-zero, leading to generalization 
of the zero-bias anomaly known from {\em dc} studies to the {\em ac} case. 
At higher frequencies clear signatures of the Coulomb 
interactions and the Kondo effect are reported. 
When properly tuned, the device can be used to generate
the spin current (with much smaller charge current still present)
and fully spin-polarized electric current.
The full frequency dependence uniquely characterizes 
the relevant low-temperature Kondo-Majorana interplay,
which we hope will stimulate and foster further
endeavors to observe it experimentally.

For experimental verification of our predictions we propose 
to use the quantum dot--Majorana mode hybrid structures, 
either based on semiconducting nanowires
\cite{Deng2016Dec} or self-organized magnetic chains
(for instance Fe atoms) deposited on superconducting
surface with side attached quantum impurities which can 
be crafted atom-by-atom 
\cite{Wiesendanger_2023}. In the first case, the experimental
spectroscopy would rely on measurement of {\em ac} ballistic 
tunneling conductance, whereas in the second situation the 
relevant measurements could be done with periodically modulated 
scanning tunneling spectroscopy. Using microwave spectroscopy 
one could also detect nonequilibrium signatures of the 
Andreev-Majorana bound states \cite{Zellekens_2022}. 
In all cases, the probed frequencies should be safely 
inside the topological gap (fractions of meV).

\begin{acknowledgements}
This research project has been supported by the National Science Centre (Poland) 
through Grant No.~2018/29/B/ST3/00937. T.D.~and I.W.~also acknowledge National 
Science Centre (Poland) Grant No.~2022/04/Y/ST3/00061.
\end{acknowledgements}

\appendix
\section{Alternative experimental realization}
\label{sec:2QD}

\begin{figure}[t!]
\includegraphics[width=0.9\linewidth]{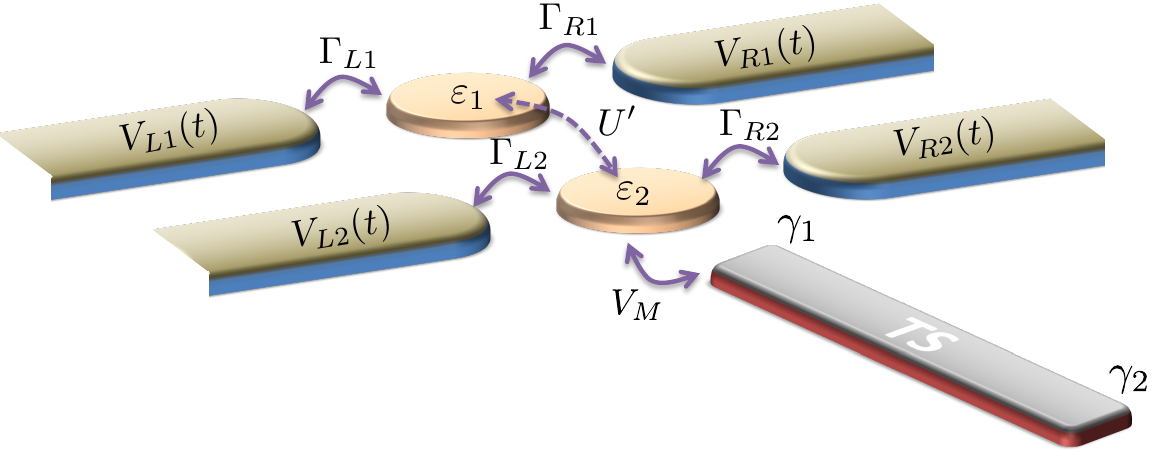}
\caption{Schematic for alternative realization of the system
	based on a double quantum dot setup.}
\label{fig:system2}
\end{figure}

As we have shown in the main text of the paper, 
application of the symmetric or antisymmetric voltage bias between 
the leads may be used to address the chosen 
response function summed over source spins.
However, even if spin-sensitive measurements are
performed, the even contribution is always a sum
of two terms, cf.~\eq{Ge}.
Separately addressing one of them would require 
application of the bias to only one spin species 
in the normal leads, which can pose a considerable experimental challenge.
However, an alternative system might be proposed, 
where $g^{\rm e}_{\s\sbar}(\w)$ would not require 
such a sophisticated biasing, at the expense of
adding one more quantum dot to the system, 
as depicted in \fig{system2}. The device presented
there has the same low energy structure, when one 
assumes that quantum dots are very small
(thus the corresponding $U$ is very large) and are placed in a strong magnetic field, 
rendering one of the spins (say spin-$\up$) irrelevant
for the low-energy processes.
Proximity of quantum dots enforces also inter-dot Coulomb repulsion $U'$,
and we assume that the direct hopping between the dots is suppressed.
In such a scenario, the spin becomes irrelevant,
while the quantum dot index plays the role of an isospin.
Only the second dot is directly coupled to the topological 
superconductor, and each quantum dot has its own bath of 
free electrons in leads $1$ and $2$, correspondingly. 
Together they constitute a single effective electrode 
possessing the isospin index, and $U'$ leads to Coulomb 
blockade and the (isospin) Kondo effect when the two QDs 
are (in total) singly occupied. Crucially, now the bias
applied to only one (iso)spin channel is easily realizable,
simply as a bias applied only to leads connected to the 
relevant quantum dot, and the response can also be measured 
in each of these separately.

\section{The case of conventional superconductor}
\label{sec:SC}

One of the most prominent results of our study is that 
the even contribution to the frequency-dependent conductance maintains the
universal value of $G_0$ over an extended range of frequencies.
Here we would like to stress, that while the existence of non-zero conductance at low $\w$ is 
a result of superconductivity, its universal character 
is a clear hallmark of a topological protection. Namely, 
we have verified through direct calculations that when 
the TS is replaced by a conventional superconductor, 
$G^{\rm e}(\w\to 0) > 0$, but is not universal. Its
magnitude becomes significant only for the quantum dot-superconductor 
coupling strength $\Gamma_{\rm S}$ exceeding $\sim U/4$.
Therefore, any effect stemming from direct coupling of 
the dot to the superconducting shell, covering the nanowire
in typical realizations of TS, can be safely ignored.

\end{document}